\documentclass[12pt, letterpaper]{JHEP3}
\usepackage{amssymb, amsmath, amsopn, amsthm}
\usepackage{epsfig}
\usepackage{psfrag}
\usepackage{subfigure}
\usepackage{cite}
\usepackage{epstopdf}
\newcommand{\eref}[1]{(\ref{#1})}

\newcommand{\cref}[1]{Chapter~\ref{#1}}
\newcommand{\beq}{\begin{equation}}
\newcommand{\eeq}{\end{equation}}
\newcommand{\ba}{\begin{array}}
\newcommand{\ea}{\end{array}}
\newcommand{\bcenter}{\begin{center}}
\newcommand{\ecenter}{\end{center}}

\def\IB{\relax\hbox{$\inbar\kern-.3em{\rm B}$}}
\def\IC{\relax\hbox{$\inbar\kern-.3em{\rm C}$}}
\def\ID{\relax\hbox{$\inbar\kern-.3em{\rm D}$}}
\def\IE{\relax\hbox{$\inbar\kern-.3em{\rm E}$}}
\def\IF{\relax\hbox{$\inbar\kern-.3em{\rm F}$}}
\def\IG{\relax\hbox{$\inbar\kern-.3em{\rm G}$}}
\def\IGa{\relax\hbox{${\rm I}\kern-.18em\Gamma$}}
\def\IH{\relax{\rm I\kern-.18em H}}
\def\IK{\relax{\rm I\kern-.18em K}}
\def\IL{\relax{\rm I\kern-.18em L}}
\def\IP{\relax{\rm I\kern-.18em P}}
\def\IR{\relax{\rm I\kern-.18em R}}
\def\IZ{\relax\ifmmode\mathchoice
{\hbox{\cmss Z\kern-.4em Z}}{\hbox{\cmss Z\kern-.4em Z}}
{\lower.9pt\hbox{\cmsss Z\kern-.4em Z}}
{\lower1.2pt\hbox{\cmsss Z\kern-.4em Z}}\else{\cmss Z\kern-.4em Z}\fi}
\def\II{\relax{\rm I\kern-.18em I}}

\def\sCC{{\kern 0.27em\vrule height1.45ex width0.03em depth0em
          \kern-0.30em\rm C}}
\def\C{{\mathchoice
  {\sCC}
  {\sCC}
  {\kern 0.225em \vrule height1.05ex width0.025em depth0em \kern-0.25em \rm C}
  {\kern 0.180em \vrule height0.78ex width0.02em depth0em \kern-0.2em \rm C}
        }}
\def\sHH{{\rm I\kern-.16em{}H}}
\def\H{{\mathchoice
  {\sHH}
  {\sHH}
  {\rm I\kern-.13em{}H}
  {\rm I\kern-.13em{}H} }}
\def\sNN{{\rm I\kern-.16em{}N}}
\def\N{{\mathchoice
  {\sNN}
  {\sNN}
  {\rm I\kern-.12em{}N}
  {\rm I\kern-.10em{}N} }}
\def\sPP{{\rm I\kern-.16em{}P}}
\def\P{{\mathchoice
  {\sPP}
  {\sPP}
  {\rm I\kern-.12em{}P}
  {\rm I\kern-.10em{}P} }}
\def\sQQ{{\kern 0.27em \vrule height1.45ex width0.03em depth0em
          \kern-0.30em \rm Q}}
\def\Q{{\mathchoice
        {\sQQ}
        {\sQQ}
  {\kern 0.225em \vrule height1.05ex width0.025em depth0em \kern-0.25em \rm Q}
  {\kern 0.180em \vrule height0.78ex width0.020em depth0em \kern-0.20em \rm Q}
        }}
\def\sRR{{\rm I\kern-0.16em{}R}}
\def\R{{\mathchoice
  {\sRR}
  {\sRR}
  {\rm I\kern-0.12em{}R}
  {\rm I\kern-0.10em{}R} }}
\def\sZZ{{\rm Z\kern-0.32em{}Z}}
\def\Z{{\mathchoice
  {\sZZ}
  {\sZZ} 
  {\rm Z\kern-0.3em{}Z}     %.3
  {\rm Z\kern-0.25em{}Z} }}  %.25
\def\ZZZ{{\rm Z\kern-0.24em{}Z}}
\def\sII{{\rm I\kern-0.16em{}I}}
\def\I{{\mathchoice
  {\sII}
  {\sII}
  {\rm I\kern-0.12em{}I}
  {\rm I\kern-0.10em{}I} }}

\def\Tr{{\rm Tr}}

\def\inbar{\,\vrule height1.5ex width.4pt depth0pt}
\font\cmss=cmss10 \font\cmsss=cmss10 at 7pt

\def\smiley{\hbox{\large$\bigcirc$\hspace{-0.80em}\raise.2ex
\hbox{$\cdot\cdot$}\kern-.61em\lower.2ex\hbox{\scriptsize$\smile$}}\ }
\def\frowny{\hbox{\large$\bigcirc$\hspace{-0.80em}\raise.2ex
\hbox{$\cdot\cdot$}\kern-.635em\lower.2ex\hbox{\scriptsize$\frown$}}\ }

\def\I{{\rlap{1} \hskip 1.6pt \hbox{1}}}

\makeatletter
\let\hangafter\@hangfrom
\makeatother

\newcommand{\drawsquare}[2]{\hbox{%
\rule{#2pt}{#1pt}\hskip-#2pt%  left vertical
\rule{#1pt}{#2pt}\hskip-#1pt%  lower horizontal
\rule[#1pt]{#1pt}{#2pt}}\rule[#1pt]{#2pt}{#2pt}\hskip-#2pt%  upper horizontal
\rule{#2pt}{#1pt}}% right vertical

\newcommand{\beqa}{\begin{eqnarray}}
\newcommand{\eeqa}{\end{eqnarray}}
\newcommand{\be}{\begin{equation}}
\newcommand{\ee}{\end{equation}}
\newcommand{\bea}{\begin{eqnarray}}
\newcommand{\eea}{\end{eqnarray}}
\newcommand{\bA}{\begin{array}}
\newcommand{\eA}{\end{array}}
\newcommand{\bc}{\begin{center}}
\newcommand{\ec}{\end{center}}

\renewcommand{\thepage}{\arabic{page}} 
\newcommand{\Yfund}{\raisebox{-.5pt}{\drawsquare{6.5}{0.4}}}%  fund

\def\be{\begin{equation}}
\def\ee{\end{equation}}
\def\bea{\begin{eqnarray}}
\def\eea{\end{eqnarray}}

\def\l{\left}
\def\r{\right}

\title{Cascades with Adjoint Matter: Adjoint Transitions}
\author{~{\normalsize \bfseries \sffamily Du\v san Simi\' c${}^{1,2}$}

~\\${}^1$Department of Physics, Stanford University\\
Stanford, CA 94305 USA \\ \vspace{0.3cm}

${}^2$ Theory Group, SLAC National Accelerator Laboratory\\
Menlo Park, CA 94025 USA \\ \vspace{0.3cm}

\email{simic@stanford.edu}\\
}

\newcommand{\bbea}{\begin{equation} \begin{aligned}} \newcommand{\eeea}{\end{aligned} \end{equation}}

\abstract{A large class of duality cascades based on quivers arising from non-isolated singularities enjoy adjoint transitions - a phenomenon which occurs when the gauge coupling of a node possessing adjoint matter is driven to strong coupling in a manner resulting in a reduction of rank in the non-Abelian part of the gauge group and a subsequent flow to weaker coupling. We describe adjoint transitions in a simple family of cascades based on a $\mathbb{Z}_2$-orbifold of the conifold using field theory. We show that they are dual to Higgsing and produce varying numbers of $U(1)$ factors, moduli, and monopoles in a manner which we calculate. This realizes a large family of cascades which proceed through Seiberg duality and Higgsing. We briefly describe the supergravity limit of our analysis, as well as a prescription for treating more general theories. A special role is played by ${\cal N}=2$ SQCD. Our results suggest that additional light fields are typically generated when UV completing certain constructions of spontaneous supersymmetry breaking into cascades - potentially leading to instabilities.}
\def\be{\begin{equation}}
\def\ee{\end{equation}}
\def\bea{\begin{eqnarray}}
\def\eea{\end{eqnarray}}

\newcommand{\cF}{\mathcal{F}}
\newcommand{\cG}{\mathcal{G}}

\newcommand{\cJ}{\mathcal{J}}
\newcommand{\cK}{\mathcal{K}}

\newcommand{\cN}{\mathcal{N}}
\newcommand{\cO}{\mathcal{O}}

\newcommand{\cS}{\mathcal{S}}

\newcommand{\cW}{\mathcal{W}}

\newcommand{\cZ}{\mathcal{Z}}

\renewcommand{\Im}{{\rm Im}\,}

\begin{document}

%===================================
\section{Introduction}
%===================================

Quiver gauge theories possessing $\cN=1$ supersymmetry arising from fractional brane configurations at non-isolated singularities (QNISs) provide a potentially large class of examples realizing spontaneous supersymmetry breaking in string theory\cite{AFGM, BMV,ABFK1,ABFK2, AK}. It is commonly believed that these examples may be embedded into duality cascades, which by the AdS/CFT correspondence would be dual to warped throats with spontaneous supersymmetry breaking deep in the interior. This is desirable as supersymmetry breaking throats are ubiquitous in a variety of string phenomenological applications to cosmology and particle physics \cite{KKLT,Gherghetta:2010cj,GP,RS,GP2,BDFKSV,MSS}. 

To realize such a construction, an improved understanding of the duality cascades arising from $\cN=1$ QNISs is needed. A challenge is posed by the presence of adjoint matter. QNISs typically contain one or more nodes whose field content is formally that of an $\cN=2$ theory, and the associated adjoints have $\cN=2$ couplings to the rest of the theory. Thus, while some of the steps in the cascade can be understood in terms of Seiberg duality, the steps involving strongly coupled adjoint nodes will involve some other dynamics, referred to as an 'adjoint transition'. Our purpose here is to understand the adjoint transitions and plot out the zoo of possible cascade-like flows using the simplest case of a certain $\mathbb{Z}_2$ orbifold of the conifold as a prototype. We expect that this will help clarify questions related to the existence of meta-stable supersymmetry breaking configurations at the bottom of such cascades. 

Cascades enjoying adjoint transitions have been studied in detail for $\cN=2$ QNISs both in supergravity and in field theory\cite{BBCC, mateo, polchinski, petrini, ofer, cremo}. In these cases the adjoint transitions proceed through Higgsing. There is a large moduli space of possible transitions, resulting in reductions in the rank of the non-Abelian part of the gauge group in varying degrees. It is quite plausible that similar dynamics persists in $\cN=1$ QNISs. This has been noted in \cite{BBCC}. Here we explore cascades with adjoint transitions in $\cN=1$ theories using the techniques of supersymmetric field theory, in the simplest case of cascades based on a $\mathbb{Z}_2$-orbifolded conifold.

To begin, we find a family of cascade-like flows which holomorphically interpolate between regimes where field theory techniques are expected to be useful and where supergravity is expected to be useful. We construct these flows as relevant deformations of a family of conformal field theories. The different flows are holomorphically connected by moving along the ultraviolet fixed-manifold. By the standard lore that superymmetric field theories undergo no phase transitions, we may thus learn about either regime by studying the field theory regime\cite{IS, SW}. In particular this should be sufficient to uncover the field theory mechanism behind the adjoint transition in either regime, and give information on the variety of possible infrared effective field theories. The definition of this family of flows is the subject of section 2. 

In section 3 we describe a discrete family of fixed points and establish a few results which are needed in section 4. 

In section 4 we study flows in the field theory regime. The flows in the field theory regime spend most of their time hovering near fixed points, with shorter flows in between which transition between the fixed points. The transitions gradually reduce the number of degrees of freedom available to the non-Abelian sector of the theory. This is similar in spirit to the original analysis due to Strassler of the conifold model \cite{S}. There the transitions between fixed points could be described using Seiberg duality. In the flows we study only half of the transitions can be understood in this manner. The other half occur on segments of the flow where there is a seemingly indefinite growth of a coupling associated to an adjoint node. These will correspond to adjoint transitions. We find that the growth of the coupling is regulated, as it is in $\cN=2$ SQCD on its Coulomb branch\cite{APS}, by a forced spontaneous breaking of the gauge group down to an infrared free or conformal subgroup. The adjoint transitions correspond to choosing a vacuum on a Coulomb branch in a manner which preserves some amount of non-Abelian gauge symmetry. In fact, the properties of such vacua are directly controlled by the physics of $\cN=2$ SQCD in a manner which we describe. After the adjoint transition, the theory is pushed back to a known interacting fixed point, however with a non-trivial number of $U(1)$ factors, moduli, and in certain cases monopoles present, forming a decoupled sector of the theory (of course, the monopoles are only massless for specially tuned values of the moduli). Thus we find a zoo of possible cascading renormalization group flows which proceed through a combination of Seiberg duality and Higgsing, producing a number of deconfined degrees of freedom along the way. 

In section 5 we discuss the supergravity limit of our analysis. Using what is known of the gauge-gravity dictionary we construct a qualitative picture of the resulting supergravity background from the analytic continuation of the field theory regime. We find a warped throat containing a sequence of singular points at which explicit fractional brane sources are present, as in \cite{ABC}. These are required to realize the U(1) factors and moduli predicted from field theory at each Higgsing event. The resulting picture  agrees well with previous work \cite{ABC}.

In section 6 we discuss the generalization of our results to other quivers. Just as understanding the mechanism behind cascades at the simplest example of an isolated singularity (the conifold\cite{KS}) appears to be sufficient in understanding the mechanism behind cascades arising at a variety of isolated singularities\cite{FHU}, we hope that the understanding we have obtained in the specific example of the orbifolded conifold is sufficient for understanding the mechanism behind cascades arising at a large class of non-isolated singularities. For these theories, we suggest a prescription for tracking the field theory in the supergravity regime analogous to the prescription applied to cascades proceeding through Seiberg duality\cite{FHU}.

We present some conclusions in section 7.

\vline
\newpage

\noindent {\bf  Table 1:}{\it ~A table of symbols}: 

\begin{equation}
\begin{array}{c|c}

\cF_N & ~five~complex ~dimensional~ family ~of~ CFTs~ described ~in ~\S2. \\
\cS_N & ~three~complex ~dimensional~ family ~of~ CFTs~ described ~in ~\S2. \\
  P_i & ~denotes~an~integer~which~is~positive~or~zero.\\
  M_i & denotes~an~integer~in~general~of~either~sign.\\
S_{N,M_2,M_4} &~ a ~member~ of~ the ~discrete~ family~ of~ CFTs ~described ~in ~\S3.\\ 
 F_{l,s} &~U(1)^l~\cN=2~gauge~theory~with~s ~massless~charged ~hypermultiplets.  \\
 \tau_i &~  \tau_i = 4\pi i/g_i^2+\theta_i/2\pi, ~ the ~holomorphic~gauge~coupling~of ~node~ i.\\
 \lambda_i &~'t~Hooft~coupling~of~node~i. \\
 \Lambda_i & ~strong~coupling~scale~of ~node ~i.\\
\end{array} \nonumber
\label{symbols} 
\end{equation}

\newpage
%=================
\section{A family of conformal field theories}
%=================

The quiver gauge theories that we consider arise by placing $N$ D3 branes at the origin of a certain $\mathbb{Z}_2$-orbifold of the conifold, which we may write as a hypersurface in $\mathbb{C}^4$:

\be
\cZ := \{ (x_1,x_2,x_3,x_4) \in \mathbb{C}^4 ~ | ~ x_1x_2 - (x_3x_4)^2 = 0 \}.
\label{eq}
\ee

This gives rise to a five-complex dimensional family of conformal field theories $\cF_N$ with gauge group $\cG$ and matter content \cite{U}:\footnote{In order to see the adjoints one must be in the appropriate duality frame.}

 \begin{equation}
\begin{array}{c|cccc}
 {\cal G}   & U(N)_1 & U(N)_2 &  U(N)_3 & U(N)_4   \\ \hline
\Phi_{22} &  1 & \rm adj & 1 & 1  \\
\Phi_{44} &  1 & 1  & 1 &  \rm adj  \\
  X_{12}, X_{21}  &  \Yfund, \overline \Yfund  & \overline \Yfund, \Yfund  & 1 & 1\\
 X_{23}, X_{32}    & 1 &  \Yfund , \overline \Yfund   & \overline  \Yfund, \Yfund   & 1 \\
 X_{34}, X_{43}  &1 &  1& \Yfund, \overline \Yfund &  \overline \Yfund, \Yfund     \\
 X_{41}, X_{14}  &\overline \Yfund, \Yfund  &  1& 1& \Yfund, \overline \Yfund      \\
\end{array}
\label{OC}
\end{equation}

The exactly marginal couplings of the conformal field theory map to the unobstructed deformations of the IIB string background.

Throughout this work, we consider only a three-complex dimensional subspace of $\cF_N$, which we denote by $\cS_N$. The subspace $\cS_N$ will be characterized by having an additional symmetry, which will help simplify our analysis. Restricting to this subspace will suffice for studying a variety of cascades enjoying adjoint transitions, as well as interpolating between supergravity and field theory regimes (to be defined more precisely below). 

\vspace{4mm}

\noindent {\it A three-complex dimensional family of fixed points}

\vspace{3mm}

Consider a family of effective field theories which are nearly free at some ultra-violet scale $\mu$ with gauge group and matter content as in \eref{OC} and superpotential:

\begin{eqnarray}
\cW = &&   h_1 \l( \Phi_{22} X_{21} X_{12} - \Phi_{22} X_{23} X_{32}  \r)  + h_2 \l(    \Phi_{44} X_{41}X_{14} -  \Phi_{44} X_{43}X_{34}  \r) \nonumber \\ 
		&& +(\eta/\mu) \l(  X_{23} X_{34} X_{43}X_{32} -X_{12}X_{21} X_{14} X_{41} \r) .
\label{free}
\end{eqnarray}
The parameters $h_1,h_2, \eta$ as well as the holomorphic gauge couplings $\tau_i$ (defined in table 1), evaluated at $\mu$, are tunable parameters. 

For appropriate values of $\{\tau_1,\tau_3 \}$, this theory is invariant under a transformation $\cJ_-$ which acts as the product of a discrete R-transformation and an outer automorphism of the gauge group, where the R-transformation acts multiplicatively with a factor of $e^{i\pi/2}$ on the gauginos and chiral superspace coordinates while keeping fixed the bottom component of all chiral matter superfields, and the outer automorphism exchanges the first and third gauge groups in \eref{OC}. 

Classically the holomorphic gauge couplings $\tau_1, \tau_3$ are simply exchanged, and thus the classical condition for $\cJ_-$ invariance is simply $ \tau_1 =  \tau_3$. However, due to an anomaly in the discrete R-symmetry, we have at the quantum level:

\be
\cJ_- : (\Lambda_1^N, \Lambda_3^N) \rightarrow  (-)^N (\Lambda_3^N, \Lambda_1^N) .
\label{J-}
\ee
Thus the action of $\cJ_-$ is only trivial if $\Lambda_1^N = (-)^N \Lambda_3^N$. We henceforth restrict to this subspace. 

In the infrared, we assume the theory flows to a non-trivial fixed point. Imposing the vanishing of the beta-functions results in non-trivial constraints on the anomalous dimensions, and hence the ultra-violet parameters. We have:

\begin{eqnarray}
\gamma_{\Phi_{44}} + 2 \gamma_{X_{14}} &&= 0 \nonumber \\ 
\gamma_{\Phi_{22}} + 2 \gamma_{X_{12}} &&=0 \nonumber  \\
1 + 2\gamma_{X_{12}} + 2\gamma_{X_{14}} &&= 0 
\label{anom}
\end{eqnarray}
where $\gamma_{\psi}$ denotes the anomalous dimension of a given field, $\psi$. The anomalous dimensions of the other fields are related to those appearing in \eref{anom} by symmetries: 

\be
\gamma_{X_{23}} = \gamma_{X_{12}},  \ \ \gamma_{X_{34}} = \gamma_{X_{14}},  \ \  \gamma_{X_{ij}} = \gamma_{X_{ji}}.
\ee
The conditions for a fixed point thus comprise three real constraints. Up to phase redefinitions of the fields, the family of effective field theories we consider has nine real parameters. Thus, we expect to obtain a complex three-dimensional family of conformal field theories, which we define to be $\cS_N$.

Note that the elements of $\cS_N$ are acted on trivially by $\cJ_-$.

\subsection{A holomorphic family of flows}

Using $\cS_N$ we define a family of flows. This is accomplished by deforming its elements by relevant perturbations. 

$\cS_N$ is three-complex dimensional,  and we may think of it as being parameterized by three of the original six holomorphic couplings, for example $\{ \tau_2, \tau_4 , \eta\}$. We define two different regimes in this space of couplings:
\begin{eqnarray}
  &&{\rm {\bf Supergravity \ regime:  }} \quad N >>   \Im \tau_2, \ \Im \tau_4 >> 1, \quad \ \eta \sim 1, \nonumber \\ 
 &&{\rm {\bf Field \ theory \ regime: } } \quad \  \infty \ >  \ \Im \tau_2, \ \Im \tau_4  >> N,  \ \ \eta << 1. 
 \label{regimes}
\end{eqnarray}
(Note that although in the field theory regime $\{ \tau_2, \tau_4 , \eta\}$ are perturbative, the other couplings $\{ \tau_1,\tau_3, h_1,h_2 \}$ in general are not, consistent with the fact  that the theories in the field theory regime continue to have $O(1)$ anomalous dimensions.)

Since the family of conformal field theories denoted by $\cS_N$ holomorphically interpolate between these two regimes, we sometimes refer to its elements as 'interpolating' conformal field theories. 

$\cS_N$ gives rise to a family of renormalization group flows by turning on vacuum expectation values for the adjoint fields. Each element in $\cS_N$ is arranged to result in a rank $(N,N-P_2,N,N-P_4)$ theory at some common scale $\nu$, for some $P_2,P_4 > 0$.\footnote{Henceforth $P_i$ will always denote non-negative integers.} Furthermore we take $P_2, P_4 << N$, as we are interested in discussing cascades. 

A flow is said to be in the field theory regime, supergravity regime or neither according to the element in $\cS_N$ from which it arises.

We note that these flows are also acted on trivially  by $\cJ_-$. This will help control the non-perturbative corrections, as will be discussed in later sections.

%=====================
\section{A discrete family of fixed points}
%=====================

Here we describe a discrete family of fixed points which will play an important role in the discussion of \S4. 

Consider the theory with matter content and superpotential:\footnote{Here and throughout the remainder of the text $M_i$ denote integers generally of arbitrary sign and it will often be assumed that $|M_i| << N$.}

 \begin{equation}
\begin{array}{c|cccc}
 {\cal G}   & U(N)_1 & U(N+M_2)_2 &  U(N)_3 & U(N+M_4)_4   \\ \hline
\Phi_{22} &  1 & \rm adj & 1 & 1  \\
\Phi_{44} &  1 & 1  & 1 &  \rm adj  \\
  X_{12}, X_{21}  &  \Yfund, \overline \Yfund  &  \overline \Yfund, \Yfund  &1 & 1\\
 X_{23}, X_{32}    & 1 &  \Yfund , \overline \Yfund   & \overline  \Yfund, \Yfund   & 1 \\
 X_{34}, X_{43}  &1 &  1& \Yfund, \overline \Yfund &  \overline \Yfund, \Yfund     \\
 X_{41}, X_{14}  &\overline \Yfund, \Yfund  &  1& 1& \Yfund, \overline \Yfund      \\
\end{array}
\label{OC'}
\end{equation}

\begin{eqnarray}
\cW = && h_1 (\Phi_{22} X_{21} X_{12}  -  \Phi_{22} X_{23} X_{32}  ) + h_2 ( \Phi_{44} X_{41}X_{14}    -\Phi_{44} X_{43}X_{34} ) ,
\label{supap}
\end{eqnarray}
in the limit: 

\be \Im \tau_2 = \Im \tau_4 = \infty. \label{decoupling} \ee
In this limit the gluons of nodes 2 and 4 decouple and so will be ignored in the subsequent discussion. We also arrange the parameters such that the theory have $\cJ_-$ invariance as in \S2. (See the discussion around \eref{J-} for specifics.) This symmetry guarantees that:

\be
\gamma_{12} = \gamma_{32}, \quad \gamma_{14} = \gamma_{34} .
\ee
(Note however that it is not necessarily the case that $\gamma_{12} = \gamma_{14}$. The symmetry that could guarantee this is broken for $M_2 \neq M_4$. We will consider the case $M_2 = M_4$ separately. For now we assume $M_2 \neq M_4$.) After imposing the symmetry constraints, the non-trivial conditions for a fixed point are:\footnote{These equations arise from demanding that the superpotenial be marginal, and that the NSZV beta functions\cite{LS,SV} for the gauge couplings of nodes 1 and 3 vanish.}

\begin{eqnarray} 
\gamma_{22} +2\gamma_{12} &&= 0 \nonumber \\  \gamma_{44} +2\gamma_{14} &&= 0  \nonumber \\  N(1 + 2\gamma_{12} + 2\gamma_{14})  -M_2(1-2 \gamma_{12}) -M_4 (1-2\gamma_{14}) &&= 0
\label{constraints}
\end{eqnarray}
Thus we have three constraints on four unknowns. This is not enough to determine the anomalous dimensions. We can solve this problem using the technique of a-maximization \cite{IW}. The analysis is simplified a great deal in the regime $M_i/N <<1$ which is the regime of interest to us. 

Let us label the R-charge of $X_{12}$ by $t$. The R-charges of the other fields are determined in terms of $t$ via the symmetries and constraints. The value of $t$ corresponding to the superconformal $U(1)_R$ maximizes $a(R)$,

\be
a(R) := 3 \Tr R^3 - \Tr R,
\ee 
where $R$ is the generator of the candidate superconformal $U(1)_R$, and the trace is over fermion species. The result of this maximization procedure is:

\be
t = 1/2 + \frac{5M_2+4M_4}{18N} -   \frac{(11 M_2^2 + 16 M_2 M_4 + 9 M_4^2)}{72N^2}  +  \cO(M_i^3/N^3),
\label{r}
\ee
where we have included the $\cO(M^2_i/N^2)$ term for later use in the next subsection. It will not be needed elsewhere. In the special case of $M_2 = M_4 :=M$ we may impose a symmetry that guarantees:
\be
\gamma_{12}=\gamma_{23} = \gamma_{14} = \gamma_{34} =: \gamma_X
\ee
In this case the anomalous dimensions are uniquely determined without a-maximization and have a compact expression:

\be
\gamma_{X_{ij}} = 1/2 - \frac{3N}{4(N+M)}, \quad \gamma_{\Phi_{ii}} = 1/2-\frac{3M}{2(N+M)}.
\ee
Of course this formula only holds in the region $3N/2 < 2N+2M < 3N$ where it satisfies the unitarity bounds of conformal field theory\cite{Mack:1975je}.\footnote{For completeness we list the anomalous dimensions for $M_2\neq M_4$:
\begin{eqnarray}
\gamma_{12},\gamma_{21}, \gamma_{32}, \gamma_{23} &&=  -\frac{1}{4} +\frac{5M_2+4M_4}{12N} +O(M_i^2/N^2), \nonumber \\
\gamma_{14}, \gamma_{41},\gamma_{31}, \gamma_{13} &&=   -\frac{1}{4} +\frac{5M_4+4M_2}{12N} +O(M_i^2/N^2), \nonumber \\
\gamma_{22} &&= \frac{1}{2} - \frac{5M_2+4M_4}{6N} +O(M_i^2/N^2), \nonumber \\ 
\gamma_{44} &&= \frac{1}{2} - \frac{5M_4+4M_2}{6N} +O(M_i^2/N^2),  \nonumber \\
\label{anom1}
\end{eqnarray}
where we have chosen to drop the $O(M_i^2/N^2)$ terms, as they will not be needed in the text. 
} 

For a wide range of $(N,M_2,M_4)$ we thus expect to have a fixed point. In this way we obtain a discrete family of CFTs. We denote its elements by $S_{N,M_2,M_4}$.

%===============
\subsection{Duality}
%===============

In \S4 it will be important to our derivations that we can Seiberg dualize nodes 1 and 3 independently at an $S_{N,M_2,M_4}$ fixed point, even though they are strongly mixed by the superpotential interactions. We can justify this in the following way. Consider a theory with gauge group and matter content identical to that of $S_{N,M_2,M_4}$ but with superpotential:
\begin{eqnarray}
\cW = && h_1 \Phi_{22} X_{21} X_{12}  +  h_2 \Phi_{44} X_{41}X_{14}   
\label{}
\end{eqnarray}
and vanishing $\lambda_{2,4}$. (See table 1 for the definition of $\lambda_i$.) In this case nodes 1 and 3 are completely decoupled. Node 3 is ordinary $\cN=1$ SQCD. We take it to sit at its Seiberg fixed point \cite{Seiberg}. It enjoys the usual Seiberg duality. On the other hand, node 1 is similar to magnetic SQCD but with the off-diagonal mesons "$\Phi_{24}, \Phi_{42}$" deleted. We refer to it as the magnetic SQCD-like theory. It is quite plausible that this theory harbors a fixed point. The conditions for this are:
\begin{eqnarray} 
\gamma_{22} +2\gamma_{12} &&= 0 \nonumber \\  \gamma_{44} +2\gamma_{14} &&= 0  \nonumber \\  N(1 + 2\gamma_{12} + 2\gamma_{14})  -M_2(1-2 \gamma_{12}) -M_4 (1-2\gamma_{14}) &&= 0
\label{constraints2}
\end{eqnarray}
(Note that because of the unequal ranks of nodes 2 and 4 there is no reason for $\gamma_{12}$ and $\gamma_{14}$ to be equal.) As in the previous section we have three constraints on four unknowns and we must use a-maximization to pin down the anomalous dimensions. If we label the R-charge of $X_{12}$ by $t$ we have:

\be
t=  1/2 +  \frac{11M_2+7M_4}{36N}  -\frac{ 17 M_2^2 + 14 M_2 M_4 + 5 M_4^2}{72N^2} + \cO(M_i^3/N^3). 
\ee
The R-charges of the other fields at node 1 are determined in terms of $t$ via \eref{constraints2}. For $M_2=M_4=:M$ there is a compact expression for the anomalous dimensions of all of the fields:\footnote{For completeness we list the anomalous dimensions for $M_2\neq M_4$:
\begin{eqnarray}
\gamma_{12},\gamma_{21} &&=  -\frac{1}{4} +\frac{11M_2+7M_4}{24N} +O(M_i^2/N^2), \nonumber \\
\gamma_{14}, \gamma_{41} &&=   -\frac{1}{4} +\frac{11M_4+7M_2}{24N} +O(M_i^2/N^2), \nonumber \\
\gamma_{22} &&= \frac{1}{2} - \frac{11M_2+7M_4}{12N} +O(M_i^2/N^2), \nonumber \\ 
\gamma_{44} &&= \frac{1}{2} - \frac{11M_4+7M_2}{12N} +O(M_i^2/N^2),  \nonumber \\
\gamma_{34}, \gamma_{43}, \gamma_{32},\gamma_{23}  &&=-\frac{1}{4} +\frac{3(M_2+M_4)}{8N} +O(M_i^2/N^2),
\end{eqnarray}
where we have chosen to drop the $O(M_i^2/N^2)$ terms, as they will not be needed in the text. 
}

\be
\gamma_{X_{ij}} = 1/2 - \frac{3N}{4(N+M)}, \quad \gamma_{\Phi_{ii}} = 1/2-\frac{3M}{2(N+M)}.
\ee
where we require $3N/2 < 2N+2M < 3N$ in order to be consistent with the unitarity bounds of conformal field theory on the dimensions of gauge-invariant operators.

For $M_2\neq M_4$ we can obtain the $S_{N,M_2,M_4}$ fixed points of the previous section via RG flow from the fixed points considered here. This is achieved by perturbing the theory by the superpotential interactions:

\be
\delta W =  -h_1' \Phi_{22} X_{23} X_{32}  - h_2' \Phi_{44} X_{43}X_{34}
\label{def}
\ee
For $M_2 \neq M_4$ one of these two couplings is always relevant and thus induces a non-trivial RG flow.\footnote{The operator dimensions are:
\be \Delta_{_{\Phi_{22}X_{23}X_{32}}} = 3 +\frac{(M_4-M_2)}{6N} +O(M_i^2/N^2), \quad \Delta_{_{\Phi_{44}X_{43}X_{34}}} = 3 -\frac{(M_4-M_2)}{6N} +O(M_i^2/N^2) .\ee } Because no accidental $U(1)$s are generated along the hypothetical flow between this fixed point and the $S_{N,M_2,M_4}$ fixed point, it must be the case that the value of the anomaly coefficient $a$ decreases along the flow \cite{IW} - yielding a consistency check on the existence of such a flow.\footnote{When accidental symmetries are generated along the flow it may in principle be possible to construct a counter-example to the hypothesis that $a_{\rm IR} <a_{\rm UV}$ always\cite{IW,ST}.} We can check that this is indeed the case. A straightforward computation yields:

\begin{eqnarray}
a_{\rm IR} &&= N^2 \l( 1 + \frac{4}{N} (M_2+M_4) - \frac{63(M_2+M_4)^2 - (M_2-M_4)^2}{36N^2}  +  \cO(M_i^3/N^3) \r ), \nonumber \\
a_{ \rm UV } && = N^2 \l ( 1 + \frac{4}{N}(M_2+M_4) -  \frac{63(M_2+M_4)^2 - 2(M_2-M_4)^2}{36N^2}  + \cO(M_i^3/N^3)  \r).  \nonumber \\  
\end{eqnarray}
Thus,
\be
 a_{\rm IR} = a_{\rm UV}- N^2 \l(    \frac{1}{36N^2} (M_2 - M_4)^2  +  \cO(M_i^3/N^3)   \r) 
\ee
and so indeed $a_{\rm IR} < a_{\rm UV}$. 

Now we come to the main point. The UV fixed point enjoys Seiberg duality acting on node 3. It is natural to conjecture that Seiberg duality holds on node 1 as well. Since this duality would be exact in the UV, it would be exact everywhere along the subsequent flow, including at the bottom.\footnote{Exact Seiberg duality along RG flows appears, for example, in the Klebanov-Strassler model\cite{S}.} Thus, assuming that Seiberg duality holds in the UV at node 1, then Seiberg duality holds independently on the two nodes of the $S_{N,M_2,M_4}$ fixed point.

Similar reasoning holds in the case $M_2= M_4$. In this case, using the techniques of Leigh and Strassler we see that this fixed point is part of a larger fixed line \cite{LS}. This fixed line includes the $S_{N,M_2,M_4}$ theory. If Seiberg duality applies node-wise at one point on this fixed line it will hold node-wise on any other point on the fixed line, in particular at the point corresponding to $S_{N,M_2,M_4}$. 

Thus all we are left to prove is that Seiberg duality holds for the magnetic SQCD-like theory at node 1. To do so consider magnetic SQCD at its fixed point. It has superpotential:

\be
\cW =  \Phi_{22} X_{21} X_{12}  +  \Phi_{24}X_{41}X_{12} +  \Phi_{42}X_{21}X_{14} +  \Phi_{44} X_{41}X_{14}  . 
\label{}
\ee
Consider adding a free sector consisting of singlets $M_{24}$ and $M_{42}$. The theory consisting of both sectors trivially enjoys Seiberg duality with the duality acting trivially on the free sector. Now consider perturbing the fixed point by the following relevant superpotential interactions:

\be
\delta W =  M_{24}\Phi_{42} + M_{42}\Phi_{24}.
\ee
Upon integrating out the massive modes we are left exactly with the magnetic SQCD-like theory. On the other hand, the UV fixed point enjoys an exact Seiberg duality as it is ordinary magnetic SQCD coupled to a free sector. Because the duality is exact in the UV, it is exact everywhere along the flow and in particular at the bottom. Thus the magnetic SQCD-like theory enjoys Seiberg duality. This completes the proof that we are free to act with Seiberg duality independently on nodes 1 and 3 of the $S_{N,M_2,M_4}$ fixed point.

%===================
\subsection{Flows}
%==============

In this section we study perturbations of an $S_{N,M_2,M_4}$ fixed point by turning on small but non-vanishing $ \lambda_2, \lambda_4$.  We will restrict to the case $M_i>0$ where $\lambda_2,\lambda_4$ are both relevant.\footnote{In the vicinity of the fixed point, the beta functions are easily computed using the NSVZ formula and \eref{anom1}:
\be
\beta_{\lambda_{2,4}} = -\frac{\lambda_{2,4}^2 f_{2,4}}{8\pi^2} \l( \frac{3M_{2,4}}{N}  + \cO(M_i^2/N^2) \r ), \nonumber \\ 
\ee
where $f_{2,4}$ are positive but scheme dependent functions. Thus $\lambda_2,\lambda_4$ are relevant perturbations when $M_4,M_2>0$.} As we will see in \S4, this is a key ingredient in understanding the cascade in the field theory regime.

We will study the fate of the flow obtained in this way by studying the Coulomb branch of the $S_{N,M_2,M_4}$ theory in the presence of the $\lambda_{2,4}$ perturbations. It is easily seen that the Coulomb branch is not lifted (non-perturbatively or otherwise) in the presence of the $\lambda_2,\lambda_4$ perturbations. Consider a vacuum classically of the form: 

\begin{eqnarray}
\Phi_{22} &&= {\rm diag} \{ 0, \dots, 0, \phi_{1}, \dots, \phi_{M_2+k_2} \}, \quad \Phi_{44} = {\rm diag} \{ 0, \dots, 0, \phi_{1}', \dots, \phi_{M_4+k_4}' \} \nonumber \\
\label{vev}
\end{eqnarray}
for some $k_2, k_4 \geq 0$. An effective potential $ \int d^2 \theta ~\delta W_{\rm np}(\phi_i,\phi_j')$ would be odd under $\cJ_-$ and hence forbidden.\footnote{We note that in the absence of $J_-$-invariance we know of no principle which would prevent such corrections.} Thus $\phi_i,\phi_j'$ parameterize exactly flat directions, and the Coulomb branch is not lifted. %\footnote{In fact, for such less symmetric flows the functional form appears in general undetermined by the selection rules and holomorphy, allowing for seemingly arbitrary corrections involving functions of the $\phi_i, \phi'_j$. In such cases we have not succeeded in solving for the vacuum structure.}

In vacua of the form \eref{vev}, at energies below $\nu_0$, which we define to be the scale of spontaneous gauge symmetry breakdown, the gauge group, $\cG'$, and matter content are:

 \begin{equation}
\begin{array}{c|ccccc}
 {\cal G'}   & U(N)_1 & U(N-k_2)_2 &  U(N)_3 & U(N-k_4)_4 & ~~U(1)^{k_2+k_4+M_2+M_4}   \\ \hline
\tilde \Phi_{22} &  1 & \rm adj & 1 & 1 & 0\\
\tilde \Phi_{44} &  1 & 1  & 1 &  \rm adj  & 0\\
  \tilde X_{12}, \tilde X_{21}  &  \Yfund, \overline \Yfund  &  \overline \Yfund, \Yfund  &1 & 1 & 0\\
 \tilde X_{23}, \tilde X_{32}    & 1 &  \Yfund , \overline \Yfund   & \overline  \Yfund, \Yfund   & 1 & 0\\
 \tilde X_{34}, \tilde X_{43}  &1 &  1& \Yfund, \overline \Yfund &  \overline \Yfund, \Yfund    & 0 \\
 \tilde X_{41}, \tilde X_{14}  &\overline \Yfund, \Yfund  &  1& 1& \Yfund, \overline \Yfund    & 0  \\
  \delta \phi_i,     \delta \phi_j' & 1  &  1& 1& 1   & 0  \\

\end{array}
\label{OC''}
\end{equation}
with superpotential:
\begin{eqnarray}
\cW = && h_1 (\tilde \Phi_{22} \tilde X_{21} \tilde X_{12}  -  \tilde \Phi_{22} \tilde X_{23} \tilde X_{32}  ) + h_2 (\tilde  \Phi_{44} \tilde X_{41} \tilde X_{14}    - \tilde \Phi_{44} \tilde X_{43}X_{34} ).
\label{supap'}
\end{eqnarray}
Below $\nu_0$ we will continue to denote by $\lambda_i$ the 't Hooft couplings of the surviving non-Abelian component of $\cG'$. Below $\nu_0$, the beta functions for $\lambda_{2,4}$ are:

\be
\beta_{\lambda_{2,4}} = \frac{\lambda_{2,4}^2 f_{2,4}}{8\pi^2} \l( \frac{3k_{2,4}}{N}  + \cO(k_i^2/N^2) \r ). 
\ee
Thus $\lambda_2$ and $\lambda_4$ are both either infrared free or conformal. 

Subsequent flow thus results in a $S_{N,-k_2,-k_4}$ conformal field theory plus a decoupled sector which is made up of the photons and moduli. We will denote this latter sector by $F_{k_2+k_4+M_2+M_4,0}$. (See table 1 for the definition of $F_{l,s}$ for general $l$ and $s$.) This is easily seen to be true for values of the moduli such that the Higgsing occurs at scales where $\lambda_2,\lambda_4$ are still weakly coupled (since in this case, we never deviate far from the fixed point to begin with). Holomorphy then implies that it remains true for generic values of the $\phi_i,\phi_j'$, including those for which $\lambda_2,\lambda_4$ deviate significantly from small values before the spontaneous breakdown occurs.

 Thus for generic vacua the endpoint of the flow from the $S_{N,M_2,M_4}$ fixed point perturbed by $ \lambda_2, \lambda_4$ is a $S_{N,-k_2,-k_4}$ conformal field theory plus a $F_{k_2+ k_4+M_2+M_4,0}$ free sector, for some non-negative integers $k_2,k_4$.

\vspace{5mm}

\noindent {\it  Non-generic points}

\vspace{3mm}

When we say we are setting out to study the fate of the flow obtained by perturbing a $S_{N,M_2,M_4}$ fixed point by turning on small but non-vanishing $\lambda_2,\lambda_4$, we don't just have in mind perturbations at generic points in moduli space. We also have in mind non-generic points, such as the origin of moduli space, etc. Interesting things may happen at non-generic points, such as an unexpected breaking of the gauge group, a flow into a new fixed point, or the appearance of additional massless particles. All of these may be relevant to the discussion of the cascade. Thus we should understand the endpoints of the flow for non-generic values of the $\phi_i,\phi_j'$ as well. 

To understand the possibilities we will employ a trick. The trick will connect, analytically, the flow of interest (whose infrared endpoints aren't known to be computable directly) to one in which the infrared endpoints are easily computable. To this end consider the theory with matter content and gauge group as in \eref{OC'} but with superpotential:

\begin{eqnarray}
\cW = &&   \sqrt{2} \l( \Phi_{22} X_{21} X_{12} - \Phi_{22} X_{23} X_{32}  \r)  + \sqrt{2} \l(    \Phi_{44} X_{41}X_{14} -  \Phi_{44} X_{43}X_{34}  \r),
\label{free}
\end{eqnarray}
where this theory is considered to be nearly free at some ultraviolet scale $\mu'$ which we imagine to be much larger than any other scale in the problem.\footnote{We further impose $\cJ_-$ invariance, which implies $\Lambda_1^{N} = (-)^{N} \Lambda_3^{N}$, as discussed in \S2.} 

For this choice of superpotential couplings, when $\Lambda_1 = \Lambda_3 =0$, the theory has exact $\cN=2$ supersymmetry. In this case the infrared phases of the theory can be understood exactly - they are simply the infrared phases of two copies of $\cN=2$ SQCD. The vacuum structure of this theory was solved in \cite{APS}. 

On the other hand, when $\Lambda_2 =\Lambda_4=0$ (but with non-zero $\Lambda_1,\Lambda_3$), we expect the theory to flow to a $S_{N,M_2,M_4}$ conformal field theory. The tuning of the superpotential couplings at the UV scale $\mu$ in \eref{free} won't change this as long as the fixed point is attractive in the infrared.\footnote{Under such a flow, the superpotential couplings will in general deviate away from their tuned UV value, to the value appropriate for the fixed point. (By value we mean with respect to some "canonical normalization" of the fields, after all corrections have been taken into account.)} This is an assumption we are free to assume, since it isn't known otherwise (and we see know reason why it shouldn't be true).\footnote{A similar assumption is made in \cite{S} in a somewhat different context.} We will henceforth make this assumption. It then follows that the flow we want to study (the perturbation of $S_{N,M_2,M_4}$ by $\lambda_2, \lambda_4$), and the flow in which there is an exact description in terms of two copies of $\cN=2$ SQCD, are holomorphically connected (through continuation in the $\Lambda_i$). 

We will use such a continuation to allow us to understand the endpoints of the flow of the $S_{N,M_2,M_4}$ theory perturbed by $\lambda_2,\lambda_4$ using known facts about the $\cN=2$ SQCD theory. The crucial step will be showing that a small but otherwise generic perturbation of the $\cN=2$ flow by a non-vanishing $\Lambda_1, \Lambda_3$, does not destroy our ability to calculate its infrared properties. If we succeed in this, then we can perform an analytic continuation, and cite the standard lore that supersymmetric gauge theories undergo no phase transitions \cite{IS,SW}, to deduce the infrared properties of the flow obtained by perturbing the $S_{N,M_2,M_4}$ theory by non-vanishing $\lambda_2,\lambda_4$.

We define two regimes:

 \begin{eqnarray}
&&{\bf Regime~A:} \ \  |\Lambda_{1,3}|  >> |\Lambda_2|, |\Lambda_4| \neq 0, \nonumber \\
&&{\bf Regime~B:} \ \    |\Lambda_2|, |\Lambda_4|  >>  |\Lambda_{1,3}| \neq 0.
\label{regimes}
 \end{eqnarray}

In regime A, the flows merge onto those of the $S_{N,M_2,M_4}$ fixed point perturbed by  small but non-vanishing $\lambda_2, \lambda_4$, and can be made to approximate them to arbitrary accuracy. Thus the endpoints of the flow from the $S_{N,M_2,M_4}$ fixed point perturbed by $\lambda_2, \lambda_4$ are identical to those of regime A.\footnote{With the caveat that any vevs we turn on have a scale smaller than $\Lambda_{1}$ and $\Lambda_3$.} 

On the other hand, regime A is holomorphically connected to regime B and so the endpoints of flows from the $S_{N,M_2,M_4}$ fixed point due to $\lambda_2, \lambda_4$ perturbations can be understood by studying flows in regime B.\footnote{In fact, if the endpoints of flows in a given regime are isolated fixed points, then not only will the endpoints of flows in the two regimes be identical as phases, they will also be identical as conformal field theories.}  Again, this follows from the standard lore that supersymmetric gauge theories undergo no phase transitions.

The usefulness of these observations lies in the fact that regime B is described over a wide range of scales by two copies of $\cN=2$ SQCD, weakly perturbed by the non-zero gauge couplings at nodes 1 and 3. The breaking to $\cN=1$ can only be felt at much lower scales corresponding to $\Lambda_1,\Lambda_3$. However, by the time these lower scales are reached, we expect most of the interesting dynamics associated with growth of $\lambda_2,\lambda_4$ to have run its course. 

Indeed, the infrared phases of asymptotically free $\cN=2$ SQCD can be understood in terms of spontaneous breakdown of the gauge group \cite{APS}, and the scale at which spontaneous breakdown occurs is bounded below by the strong coupling scale of the theory (in our case $\Lambda_2$ and $\Lambda_4$). The boundedness of the scale of the gauge symmetry breaking by $\Lambda_2$ and $\Lambda_4$ (and the fact that there is only trivial $\cN=2$ physics below it) thus implies a decoupling of the non-trivial $\cN=2$ physics from the $\cN=1$ physics associated with the non-vanishing of $\Lambda_1$ and $\Lambda_3$. We thus find that the infrared phases of the flows of regime B are classified by a choice of vacuum in the two copy $\cN=2$ SQCD theory. %This is a powerful observation because the vacuum structure of $\cN=2$ SQCD is known exactly \cite{APS}.

Since we are interested in understanding the non-generic points on the Coulomb branch of the $S_{N,M_2,M_4}$ theory perturbed by $\lambda_2,\lambda_4$, we will restrict the discussion to flows arising from vacua on the Coulomb branch of the two copy $\cN=2$ SQCD theory. Such vacua are classified by four non-negative integers $\{ k_2,k_4;r,r' \}$ which specify the amount of spontaneous gauge symmetry breaking and the number of monopole hypermultiplets respectively\cite{APS}.\footnote{For simplicity let us restrict to points with mutually local particles.} Let us imagine choosing such a vacuum for a flow in regime B, and let us denote the scale of spontaneous gauge symmetry breaking by $\nu_0$. 

Below $\nu_0$, the gauge group $\cG'$ and matter content for a flow arising in a general such vacuum is given by: 

 \begin{equation}
\begin{array}{c|ccccc}
 {\cal G'}   & U(N)_1 & U(N-k_2)_2 &  U(N)_3 & U(N-k_4)_4 & ~~U(1)^{k_2+k_4+M_2+M_4}   \\ \hline
\tilde \Phi_{22} &  1 & \rm adj & 1 & 1 & 0\\
\tilde \Phi_{44} &  1 & 1  & 1 &  \rm adj  & 0\\
  \tilde X_{12}, \tilde X_{21}  &  \Yfund, \overline \Yfund  &  \overline \Yfund, \Yfund  &1 & 1 & 0\\
 \tilde X_{23}, \tilde X_{32}    & 1 &  \Yfund , \overline \Yfund   & \overline  \Yfund, \Yfund   & 1 & 0\\
 \tilde X_{34}, \tilde X_{43}  &1 &  1& \Yfund, \overline \Yfund &  \overline \Yfund, \Yfund    & 0 \\
 \tilde X_{41}, \tilde X_{14}  &\overline \Yfund, \Yfund  &  1& 1& \Yfund, \overline \Yfund    & 0  \\
  \delta \phi_i,     \delta \phi_j' & 1  &  1& 1& 1   & 0  \\
e_l, \tilde e_l  & 1& 1_{q},1_{-q}& 1& 1 & Q_{il}, - Q_{il} \\ 
e_l', \tilde e_l'  & 1& 1& 1& 1_{q'}, 1_{-q'} & Q'_{jl}, -Q'_{jl} \\ 
\end{array}
\label{ngOC''}
\end{equation}

\noindent Here $Q_{il}$ and $Q'_{il}$ are the charge matrices for the monopoles under $U(1)^{k_2+k_4+M_2+M_4}$ and $q,q'$ are their charges under the $U(1)$ factors of $U(N-k_2)_2$ and $U(N-k_4)_4$ respectively. Below $\nu_0$ we will continue to denote by $\lambda_i$ the 't Hooft couplings of the surviving non-Abelian component of $\cG'$. 

It is easily seen that below $\nu_0$ $\lambda_2$ and $\lambda_4$ are infared free or conformal. In contrast, $\lambda_1$ and $\lambda_3$ continue to flow towards strong coupling. We thus expect subsequent flow to result in a $S_{N,-k_2,-k_4}$ conformal field theory plus a sector which is made up of the photons, moduli and possibly massless monopoles. This latter sector is denoted by $F_{k_2+k_4+M_2+M_4,r+r'}$.

We may ask to what extent this expectation is really fulfilled. In principle, non-perturbative corrections due $\Lambda_1,\Lambda_3$ could lift some of the monopoles.\footnote{Although there are a host of other corrections, those are plausibly absorbable into a redefinition of the couplings, and so we will ignore them.} Neglecting at first such corrections, the superpotential below $\nu_0$ is given by:

\begin{eqnarray}
\cW/\sqrt{2} = &&  \tilde \Phi_{22} \tilde X_{21} \tilde X_{12}  -  \tilde \Phi_{22} \tilde X_{23} \tilde X_{32}   + \tilde  \Phi_{44} \tilde X_{41} \tilde X_{14}    - \tilde \Phi_{44} \tilde X_{43}X_{34}  \nonumber \\ &&  +\frac{q}{N-k_2}  \Tr \{ \Phi_{22} \} \sum_{l=1}^{r}  e_l\tilde e _l  +\frac{q'}{N-k_4} \Tr \{ \tilde \Phi_{44} \} \sum_{l=1}^{r'}  e'_l\tilde e' _l  \nonumber \\ && + \sum_{il}   Q_{il}  \delta \phi_i e_l \tilde e_l +  \sum_{jl}   Q'_{jl}  \delta \phi'_j e'_l \tilde e'_l.
\label{ngsup}
\end{eqnarray}
For vanishing $\Lambda_1, \Lambda_3$, this expression is exact (due to $\cN=2$ superymmetry). For non-vanishing $\Lambda_1,\Lambda_3$ we are in principle in danger of generating a monopole bilinear:
\be
\delta W =  \sum_l m_l e_l \tilde e_l  + \sum_l m_l' e_l' \tilde e_l',
\label{mass}
\ee
where the $m_i$ may in general be allowed to depend on the moduli. However such corrections are easily seen to vanish due to an R-symmetry (and a consideration of various limits in the couplings):

\vspace{1mm}

\noindent First, assume that some of the $m_i$ are non-vanishing. Consider the R-symmetry under which the microscopic variables $\Phi_{ii}$ and $X_{ij}$ have charge $0,1$ respectively. From \eref{ngsup} it is easily deduced that the monopole fields $e_i,\tilde e_i$ have charge +1 under this transformation. Thus the $m_i$ must be neutral. However, under this R-symmetry the holomorphic strong coupling scales of nodes 2 and 4 are neutral, while those of 1 and 3 have  positive definite charge. However, this is inconsistent with the requirement that the $m_i  \rightarrow 0 $ as $\Lambda_1,\Lambda_3 \rightarrow 0$ in various ways. Thus, the $m_i$ must vanish identically.  

\vspace{2mm}

However, even if corrections of the form \eref{mass} were not forced to vanish, one could still hope to eliminate them through a shift in the moduli. This should be possible in regime B where such corrections are small compared to the scale at which the moduli first "appear", which is roughly $\nu_0$. The only potential danger is that upon continuation to regime A the required shift grows so large that the scale associated with the appearance of the monopoles becomes comparable to (or larger than) the scale at which the flows in regime A begin to approximate those of the $S_{N,M_2,M_4}$ fixed point perturbed by $\lambda_2,\lambda_4$. In such a situation it might be incorrect to conclude that the endpoint of the flow in the $S_{N,M_2,M_4}$ fixed point perturbed by $\lambda_2,\lambda_4$ has light monopoles. 

We thus find a series of plausible arguments which imply that the possible infrared endpoints of the flow from the $S_{N,M_2,M_4}$ fixed point perturbed by small but non-vanishing $\lambda_2, \lambda_4$ are the set of $S_{N,-k_2,-k_4}$ fixed points with an additional decoupled $F_{k_2+k_4+M_2+M_4,r+r'}$ sector, in such a way that is precisely determined by the underlying $\cN=2$ theory.

Henceforth we will refer to the conformal field theory consisting of an $S_{N,M_2,M_4}$ fixed point and an additional decoupled sector consisting of an $F_{l,s}$ theory evaluated at zero coupling, as an $S_{N,M_2,M_4}\times F_{l,s}$ fixed point. 

\vspace{3mm}

\noindent {\it Remarks}

\begin{itemize}
\item We note that not all vacua of the two copy $\cN=2$ SQCD theory lead to stable vacua in the $S_{N,M_2,M_4}$ theory perturbed by non-vanishing $\lambda_2,\lambda_4$. For example, a generic point on the Coulomb branch gives masses to all of the quark fields. This results in gaugino condensation at nodes 1 and 3 which in turn generates a destabilizing potential for the moduli. This is possible because the value of the condensate is a function of the quark masses which in turn are a function of the moduli. The unstable vacuum will roll around until it settles into a supersymmetric ground state. 

\item{Although the couplings in the second line of \eref{ngsup} are marginal for sometime below $\nu_0$, they become highly irrelevant near the $S_{N,-k_2,-k_4}$ fixed point due to the positive anomalous dimensions that $\Phi_{22}$ and $\Phi_{44}$ acquire.\footnote{One may worry that the monopoles acquire compensating negative anomalous dimensions. However, because the monopoles are gauge-singlets with respect to the non-Abelian component of the gauge group, their operator dimensions should satisfy the unitarity bound: \be \Delta_{e}, \Delta_{\tilde e}, \Delta_{ e'}, \Delta_{\tilde e'} \geq 1, \ee thus preventing the acquisition of a negative anomalous dimension. Near the $S_{N,-k_2,-k_4}$ fixed point we will have: \be\Delta_{\tilde \Phi_{22} e \tilde e},  \Delta_{\tilde \Phi_{44} e' \tilde e'}  = 3.5 + O(k_i/N).\ee} Thus, such terms do not spoil the conclusion that the infrared endpoint of the flow is an  $S_{N,-k_2,-k_4}\times F_{k_2+k_4+M_2+M_4,r+r'}$ fixed point.}

\item{It may be interesting to understand the importance to our discussion of the vacua of the two copy $\cN=2$ SQCD theory with mutually non-local particles.}
\end{itemize}

%=================
\section{Flows in the field theory regime}
%=================

We now turn to the problem of studying the flows in the field theory regime. (The flows that we will consider were defined at the end of \S2.) 

Consider first the simplest such flow corresponding to an element $s_0 \in \cS_N$ with $\lambda_2 = \lambda_4 = \eta = 0$. The infrared limit of this theory is the $S_{N,-P_2,-P_4}$ fixed point, with $P_i>0$. This is an element of the discrete family of fixed points defined in \S3. 

%====================
\subsection{Seiberg transitions}
%====================

Consider now flows originating not from $s_0$ but from nearby $s_0$ - from elements of $\cS_N$ with small but non-zero $\{  \lambda_2, \lambda_4 , \eta \}$. Such flows can be arranged to pass arbitrarily closely to the $S_{N,-P_2,-P_4}$ fixed point. We denote the scale at which the theory is closest to this fixed point by  $\mu_0$. 

We assume sufficiently small $\{  \lambda_2, \lambda_4 , \eta \}_{\mu_0}$ such that conformal perturbation theory is useful. The leading terms in the $\beta$-functions near the fixed point may then be computed using the anomalous dimensions derived in \S3:\footnote{The $f_i$ are scheme dependent functions of the couplings which are positive \cite{S, Seiberg, SV}. }
\begin{eqnarray}
\beta_{\lambda_{2,4}} && = \frac{\lambda_{2,4}^2 f_{2,4}}{8\pi^2} \l( \frac{3P_{2,4}}{N}  + \cO(P_i^2/N^2) \r ) \nonumber \\ 
\beta_{\eta} && = -\eta \l ( \frac{3P_2 + 3P_4}{2N} + \cO(P_i^2/N^2)   \r).
\label{step1}
\end{eqnarray}
Thus, near the $S_{N,-P_2,-P_4}$ fixed point the 't Hooft couplings $\lambda_2$ and $\lambda_4$ are irrelevant while the quartic coupling $\eta$ is relevant. The growth of $\eta$ thus drives the flow away from the fixed point and so conformal perturbation theory about the $S_{N,-P_2,-P_4}$ fixed point eventually ceases to be useful. 

In order to understand how to follow the resulting flow let us consider the case of $\cN=1$ SQCD with $3n_c/2 < n_f < 2n_c$ at its interacting fixed point. We deform the fixed point by the quartic superpotential:
\be
W = \eta~ Q\tilde Q Q \tilde Q.
\ee
Because $n_f<2n_c$ the anomalous dimensions are such that this operator is relevant\cite{S,Seiberg}. In order to understand the resulting flow we change duality frames to magnetic SQCD. The dual theory with the $\eta$-deformation has superpotential\cite{S}:
\be
W_{\rm dual} = M \tilde q q + \eta M^2.
\ee
Thus $\eta$ has been transformed into a mesonic mass term.  When the physical value of $\eta$ approximates unity we integrate out $M$, yielding: 
\be
W_{\rm dual} = -\frac{1}{4\eta}\tilde q q  \tilde q  q .
\ee
Because $n_f > 2(n_f-n_c)$ the anomalous dimensions are now such that this quartic deformation is irrelevant and so flows to zero in the IR. Thus the theory flows to the interacting fixed point of $SU(n_f - n_c)$ SQCD. The singlets are no longer present.

Thus, in our theory, in order to follow the RG flow into and past the region where $\eta$ becomes strong we switch duality frames, replacing nodes 1 and 3 by their Seiberg duals, and integrate out the massive modes.\footnote{This is examined more carefully in \S3.1.} The result is:

\begin{eqnarray}
W &&=  \tilde \Phi_{22} \tilde X_{21} \tilde X_{12} + \tilde \Phi_{44} \tilde X_{41} \tilde X_{14}  - \tilde \Phi_{22} \tilde X_{23} \tilde X_{32} - \tilde \Phi_{44} \tilde X_{43} \tilde X_{34}  \nonumber \\
&&  + \tilde \eta ~  \tilde X_{23} \tilde X_{34} \tilde X_{43} \tilde X_{32}
 - \tilde \eta ~\tilde X_{21} \tilde X_{14} \tilde X_{41} \tilde X_{12}  
  \label{aftseib}
\end{eqnarray}
with $\tilde \Phi_{22} = M_{22} = \tilde M_{22}$ and $\tilde \eta = -1/\eta$. 

The theory is back to a similar form as before except now the flow is near an $S_{N'P_4P_2}$ fixed point, with $N' = N-P_2-P_4$. As in our example with ordinary SQCD, the quartic couplings are now irrelevant and were it not for the non-zerodness of $\lambda_{2,4}$, the theory would flow to a fixed point. The beta functions are:

\begin{eqnarray}
\beta_{\lambda_{2,4}} && = -\frac{\lambda_{2,4}^2 f_{2,4}}{8\pi^2} \l( \frac{3P_{4,2}}{N'}  + \cO(P_i^2/N'^2) \r ) \nonumber \\ 
\beta_{\tilde \eta} && = \tilde \eta \l ( \frac{3P_{2}+3P_4}{2N'} + \cO(P_i^2/N'^2)   \r)
\label{last}
\end{eqnarray}
Thus, while the quartic couplings are irrelevant, the adjoint node gauge couplings are relevant, and their growth appears to drive the theory away from a known fixed point. This is the onset of an adjoint transition. 

%=======================
\subsection{Adjoint transitions}
%=======================

We denote the scale at which the flow is closest to the $S_{N'P_4P_2}$ fixed point by $\mu_1$. The theory at scales $\mu < \mu_1$ can be understood as a perturbation of this fixed point by  $\{ \lambda_2, \lambda_4, \tilde \eta\}_{\mu_1}$.

Since the quartic perturbation is initially small and irrelevant the problem of understanding the effects of the growth of $\lambda_2,\lambda_4$ reduces to an analysis of $S_{N'P_4P_2}$ perturbed by $\{ \lambda_2,\lambda_4,0\}_{\mu_1}$. This problem was studied in \S3.2 with the conclusion that as the couplings grow they induce spontaneous breakdown of the gauge symmetry, in a manner which is determined by a related $\cN=2$ theory. For example, if $\lambda_2$ is the first to reach strong coupling, then just below its strong coupling scale the theory has a description as a perturbation of a $S_{N',-k_2,P_2}$ fixed point, with an additional irrelevantly coupled $F_{P_4+k_2, r}$ sector. The possibilities for this sector were determined by the related $\cN=2$ theory. We thus find at scales below the strong coupling scale of $\lambda_2$, that the flow is taken near a $S_{N',-k_2,P_2}\times F_{P_4+k_2, r}$ fixed point.\footnote{The terminology of an $S_{N',-k_2,P_2}\times F_{P_4+k_2, r}$ fixed point is introduced at the end of \S3.2.} We refer to the dynamics which occurs during the flow between the vicinity of the $S_{N',-k_2,P_2}$ fixed point and that of the $S_{N',-k_2,P_2}\times F_{P_4+k_2, r}$ fixed point as an adjoint transition. 

The $\beta$-functions near the  $S_{N',-k_2,P_2}\times F_{P_4+k_2, r}$ fixed point read:\footnote{We suppress the infrared free beta functions of the couplings in the  $F_{P_4+k_2, r}$ sector.}

\begin{eqnarray}
\beta_{\lambda_{2}} && = \frac{\lambda_{2}^2 f_{2}}{8\pi^2} \l( \frac{3k_{2}}{N'}  + \cO(P_2^2/N'^2, k_2^2/N'^2) \r ) \nonumber \\ 
\beta_{\lambda_{4}} && = -\frac{\lambda_{4}^2 f_{4}}{8\pi^2} \l( \frac{3P_{2}}{N'}  + \cO(P_2^2/N'^2, k_2^2/N'^2) \r ) \nonumber \\ 
\beta_{\tilde \eta} && = \tilde \eta \l ( \frac{3P_{2}-3k_2}{2N'} + \cO(P_2^2/N'^2, k_2^2/N'^2)   \r).
\end{eqnarray}
Thus, after the adjoint transition $\lambda_2$ becomes an irrelevant perturbation, while $\lambda_4$ continues to be relevant. 

 If the quartic coupling remains non-relevant $(k_2 \leq P_2)$ then we may continue to ignore it. The coupling $\lambda_4$ will grow until its strong coupling scale is reached and a second adjoint transition occurs. The resulting flow drives the theory to a perturbation of $S_{N',-k_2,-k_4} \times F_{P_2+P_4+k_2+k_4, r+r'}$. 
 
 On the other hand, if the quartic coupling is relevant as a perturbation of  $S_{N',-k_2,P_2}\times F_{P_4+k_2, r}$
$(k_2 > P_2)$ and the strong coupling scale of $\lambda_4$ is sufficiently small, then it may grow to unity before the $\lambda_4$ transition occurs.  In this case a Seiberg transition occurs as described in the previous section, taking the flow near a $S_{N'+P_2-k_2,-P_2,k_2}\times F_{P_4+k_2, r}$ fixed point. Around this fixed point $\lambda_4$ is the only remaining relevant coupling, and so the second adjoint transition occurs uninterrupted. The result is to take the theory near a $S_{N'+P_2-k_2,-P_2,-P_4'} \times F_{P_4+P_4'+2k_2, r+r'}$ fixed point.

In either of these two cases the result of the flow from  $S_{N'P_4P_2}$ perturbed by  $\{ \lambda_2, \lambda_4, \tilde \eta\}_{\mu_1}$ is to induce two adjoint transitions (with the possibility of a Seiberg transition in between), one each on nodes 2 and 4, after which the flow is driven near a $S_{N'',-k_2',-k_4'} \times F_{l,s}$ fixed point for some $(N'' ,k_2',k_4', l,s)$. The precise value of $(N'' ,k_2',k_4', l,s)$ as a function of the $(N, P_2,P_4, P_4',k_2,k_4)$ above is determined by which of the two cases is realized.

%======
\subsection{Subsequent cascade steps}
%======

As we have seen in the previous subsection, the result of the adjoint transitions is to take the flow back near a  $S_{N'',-k_2',-k_4'} \times F_{l,s}$ fixed point for some $(N'' ,k_2',k_4', l,s)$. We denote the scale at which the flow is closest to this fixed point by $\mu_2$.

%\begin{figure}[h]
%\begin{center}
%\psfrag{A12}[cc][][1]{$A_{1,2}$}
%\psfrag{B12}[cc][][1]{$B_{1,2}$}
%\includegraphics[0 ,0][766,200]{cascade.png}
%\caption{\it  In the field theory regime the cascade snakes its a way between the fixed %points (represented here by the oval shaped dots). The transitions (Seiberg or adjoint) %occur in the middle, far from the dots. The ranks of the quiver are gradually depleted %as the flow moves towards the infrared (in the direction of the arrow).}
%\label{figure}
%\end{center}
%\end{figure}

\begin{figure}[h]
\begin{center}
\includegraphics[width=4in]{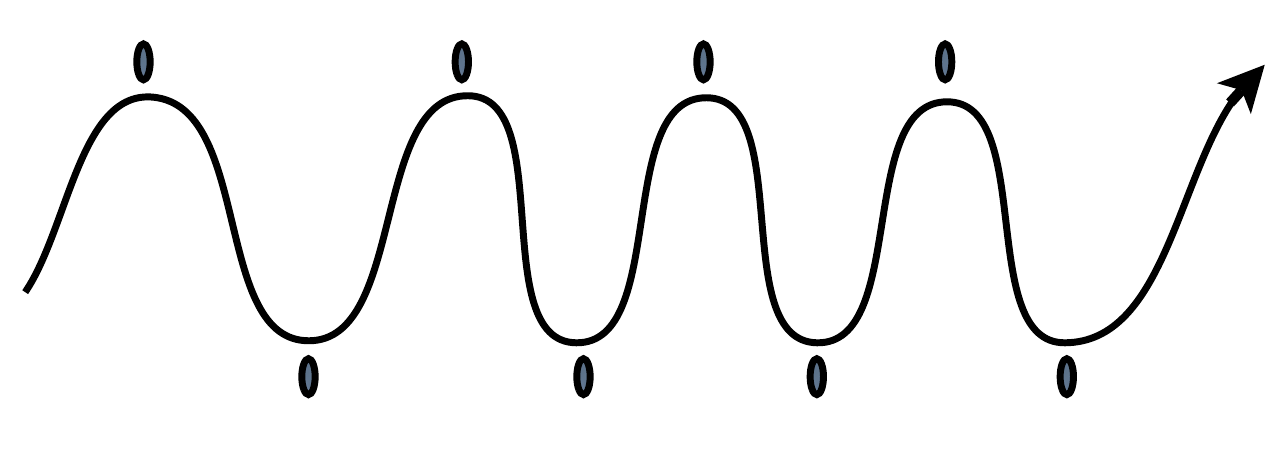}
\vspace{-2mm}
\end{center}
\noindent
\caption{\it  In the field theory regime the cascade snakes its way between the fixed points (represented here by the oval shaped dots). The transitions (Seiberg or adjoint) occur in the middle, far from the dots. The ranks of the quiver are gradually depleted as the flow moves towards the infrared (in the direction of the arrow).}
\label{throat}
%\vspace{-6mm}
\end{figure}

For $k_2' = k_4' = 0$, $\{ \lambda_2,\lambda_4, \eta\}_{\mu_2}$ are exactly marginal perturbations and the flow terminates onto a conformal field theory. In fact, up to the $ F_{l,s}$ factor, the flow terminates onto an element of $S_{N'}$. 

In the case of $k_2',k_4' >0$, around $\mu_2$ $\eta$ is relevant while $\lambda_2, \lambda_4$ are irrelevant. The situation is identical to the situation around \eref{step1} before the first Seiberg transition, except with the replacements $(N, P_2, P_4) \rightarrow (N'', k_2', k_4')$ and the additional $F_{l,s}$ factor.

As in that situation, the resulting flow due to the growing quartics is best described by performing a Seiberg duality on nodes 1 and 3. Integrating out the massive modes takes the flow back to a $S_{N''k_4'k_2'}\times F_{l,s}$ fixed point with $N''' = N'' - k_2' - k_4'$. This is identical to the situation around \eref{last} before the first adjoint transition, except with $(N', P_2,P_4) \rightarrow (N'', k_2', k_4')$ and the additional $F_{l,s}$ factor. As in that situation $\lambda_{2,4}$ grow and additional adjoint transitions occur, and so on, resulting in a cascade. This flow between fixed points is depicted in figure 1.

Thus we find a family of self-similar and cascade-like flows, which alternate between regions where the physics is close to an $S \times F$ fixed point followed by regions where $\cN=2$ type dynamics is dominant and Higgsing occurs. Thus these cascades proceed through a combination of Seiberg duality and Higgsing in an alternating manner. When the ranks of the nodes are no longer large and nearly equal some of the approximations we have been making break down and a more refined analysis is needed. 

\vspace{3mm}

\noindent {\it Remarks}

\begin{itemize}
\item In describing the RG flow, we have implicitly been assuming that a faithfull description is given by tracking  a few superpotential couplings $\{ \lambda_2,\lambda_4, \eta \}$. In reality, there are a large number of Kahler couplings generated along the RG flow which our analysis ignores. This is justified when such corrections are irrelevant. Thus, if $\delta \cK$ is any non-redundant correction to the Kahler potential, we require:
\be
\Delta_{\delta \cK} - 2 > 0. 
\label{bound}
\ee
This is easily satisfied if the theory is weakly interacting, and may be plausibly satisfied in the theories under consideration here. 

We note that the irrelevance of such corrections in interacting theories is also implicit in Strassler's analysis of the conifold, which is similar in spirit to the analysis here \cite{S}. A check on whether such corrections can be justifiably neglected is whether our conclusions are in qualitative agreement with supergravity. This is discussed in \S5. 

\end{itemize}
\subsection{$\cJ_-$ non-invariant flows}

In \S2 we chose our flows to arise as relevant deformations of a special complex co-dimension 2 subspace $\cS_N \subset \cF_N$, which was defined by the property that the transformation $\cJ_-$, defined in \S2, acted trivially on its elements. The more generic case corresponds to selecting an element from $\cF_N\setminus \cS_N$. The resulting flows are considerably more complicated as there are effectively five different couplings instead of three. 

Furthermore, in such cases, we know of no principle which forbids corrections to the Coulomb branch of the kind ruled out in \S3.2. The super-selection rules and dynamical considerations do not seem to sufficiently constrain the form of $\delta W_{\rm np}$, making it difficult to say anything about the vacuum structure. It would be interesting if such corrections could be sufficiently constrained to say something useful.

%===============
\section{The supergravity regime}
%===============

Here we discuss the supergravity limit of the above discussions. Explicit supergravity solutions for a class of flows similar to those considered here have been constructed in \cite{ABC}. 

The cascades discussed above proceeded through a combination of Seiberg duality and the spontaneous breakdown of gauge symmetry. The spontaneous breakdown events were spread along the renormalization group scale in a hierarchical manner. This hierarchy is to a certain extent holomorphic, as the breakdown events are specified in terms of vacuum expectation values of holomorphic fields. Thus we expect this hierarchy to continue analytically to a hierarchy in the supergravity regime. 

In the supergravity regime, the renormalization group scale is an emergent dimension along which physics is approximately local.  The deconfined degrees of freedom produced at each Higgsing event (the monopoles, moduli, and U(1)s) must be localized along this dimension in a manner consistent with the stretch of RG time over which they deconfine and decouple - ie: "peel off" - from the original non-Abelian degrees of freedom which source the bulk geometry\cite{PP}. 

The bulk geometry in the region where the theory is cascading and approximately conformal has the metric\cite{ABC}:

\be
ds^2  = h(r)^{-1/2}dx^2 + h(r)^{1/2}(dr^2+ (dT_{1,1}/\mathbb{Z}_2)^2)
\ee
where $T_{11}/\mathbb{Z}_{2}$ denotes a $\mathbb{Z}_2$ orbifold of the space $T_{11}$ as described in \cite{ABC} and the coordinate $r$ is related to the RG scale $\mu$ via $\mu = r/\alpha'$. 

The orbifold produces a singularity stretching along $r$ whose topology is locally $\mathbb{R}\times S^1 \cong \mathbb{C}$. Because adjoint fields are naturally identified with the motions of fractional branes along the non-isolated singularity \cite{BMV,BBCC}, it is natural to expect that the deconfined degrees of freedom are localized around specific radial locations along the singular locus of the bulk geometry. This is indeed seen in \cite{ABC}. 

In \cite{ABC} a common feature of all of the studied flows was that the adjoint transitions caused a reduction in rank of the gauge group in a manner matching the numerology of Seiberg duality. Such numerology is easily reproduced by the flows of the previous sections by choosing the Higgsing vacua appropriately. 

Our field theory analysis (continued to the supergravity regime) adds to previous results in the supergravity regime in several ways. First, it establishes (at least for $\cJ_-$ invariant flows) the presence of exactly flat moduli and the existence of special points with massless monopoles in these string backgrounds, and allows for the computation of some of their properties (some of which may be difficult to compute in supergravity). Second, our approach sheds light on the field theory mechanism behind the adjoint transitions, and allows the derivation of the precise low energy effective field theory for a large family of flows in a unified manner. This point of view makes manifest the fact that the adjoint transitions need not follow the numerology of Seiberg duality, and that the cases in which they do not are in fact more generic. 

%===============================
\section{More general quivers and a prescription}
%===============================

We have seen that the adjoint transitions are well approximated by replacing each strongly coupled adjoint node by a copy $\cN=2$ SQCD. We expect this to hold quite generally for any $\cN=1$ QNIS. 

When a cascade is in the supergravity regime, it is often still useful to have a field theory description for the renormalization group flow. For quivers based on isolated singularities, a prescription for keeping track of the field theory is to\cite{FHU, richard}:

\vspace{3mm}

$\bullet$ Compute the beta functions. 

$\bullet$ Run the inverse couplings until one approximately reaches zero. 

$\bullet$ Apply a Seiberg duality on the node whose inverse coupling approximates zero. 

$\bullet$ Integrate out massive mesons. 

$\bullet$ Recompute the beta functions for updated matter content.

$\bullet$ Repeat. 

\vspace{3mm}

\noindent When the node approaching strong coupling has adjoint matter, Seiberg duality cannot be applied. Instead our results point to the prescription:

\vspace{3mm}

 $\bullet$ Approximate the strongly coupled node by a copy of $\cN=2$ QCD. 

 $\bullet$ Choose a point on the Coulomb branch.
 
 $\bullet$ Integrate out massive modes, and update the Wilsonian effective action. 

$\bullet$ Recompute the beta functions for updated matter content.

$\bullet$ Determine next node to hit strong coupling. 

\vspace{3mm}

We have found strong evidence that this is the correct prescription for flows at the $\mathbb{Z}_2$ orbifolded conifold (at least for sufficiently symmetric flows), and it is natural to conjecture that it holds for any QNIS (at least for sufficiently symmetric flows).\footnote{We place the qualifier "for sufficiently symmetric flows" because we expect the necessity to impose a symmetry, in order to retain calculability, will continue in other quivers as well. For the flows considered here this is briefly discussed in \S4.4 and \S3.2.}

\vspace{3mm}

%========================
\section{Conclusions}
%========================

Our main interest in this paper was to understand the RG flows of $\cN=1$ quiver gauge theories with adjoint matter, coupled in an $\cN=2$ like manner to the rest of the theory. Such theories commonly arise from fractional brane configurations at non-isolated singularities and form an infinite class of theories. 

The main hinderance in understanding these theories thus far has been the adjoint matter. The field theory dynamics in the regime where the gauge coupling associated to the adjoint becomes strong has not been well understood thus far. We argued for a solution to this problem in the concrete example of the $\mathbb{Z}_2$-orbifolded conifold, finding that the dynamics is correctly approximated by replacing the sector containing the adjoint by a copy of $\cN=2$ SQCD. This is powerful because the IR structure of this theory is known exactly\cite{APS}. The resulting Higgsing produces various numbers of moduli, $U(1)$ factors, and monopoles which decouple from the remaining non-Abelian degrees of freedom in a calculable manner. Thus these cascades proceed through a combination of Seiberg duality and Higgsing. 

By mapping our results into the supergravity regime, we argued that the geometry should contain a series of regions dispersed along the radial coordinate where these deconfined states are localized. We expect the $U(1)$ factors and moduli to arise from explicit factional brane sources in the geometry as in \cite{ABC, BBCC}, while the monopoles are expected to have a more 'non-perturbative' origin in terms of tensionless wrapped D3 branes.

Our results strongly suggest that the dynamics of the adjoint in more general quivers is also faithfully reproduced by $\cN=2$ SQCD. This would open up the way for a detailed understanding of how to embed the metastable models of \cite{BMV,ABFK1,ABFK2,AK,AFGM} into duality cascades, which was the primary motivation of this work.  Via the gauge-gravity correspondence such an embedding would realize a meta-stable state inside of a warped throat geometry, and so could be of some interest. However, an interesting feature suggested by our analysis is that any to attempt to do so would yield additional light fields which survive in the IR, not present in the original models. This could affect the stabilization of these models as the scalars in these extra sectors may induce runaways. We plan to discuss this problem in forthcoming work \cite{toapp}.

\bigskip
\centerline{\bf{Acknowledgements}}

We wish to thank R. Argurio, F. Benini, M. Bertolini, S. Cremonesi, C. Closset, S. Franco, S. Kachru, N. Seiberg,  S. Shenker, M. Unsal and E. Witten for interactions which were useful in bringing the manuscript to its final form. We especially thank S. Franco, S. Kachru and the JHEP referees for their reading of the manuscript and valuable feedback. We acknowledge the hospitality of TASI, where part of this work was done. We are supported by the Mayfield Stanford Graduate Fellowship, the Stanford Institute of Theoretical Physics, and by the DOE under contract DE-AC03-76SF00515.

\bigskip

\appendix

\end{document}